\documentclass[onecolumn,preprint,a4paper,12pt,aps]{revtex4}
\usepackage[left=19mm,right=19mm,top=17mm,bottom=17mm,includehead,includefoot]{geometry}

\usepackage{amsmath,amssymb,amsfonts}
\usepackage{graphicx}
\usepackage{here}
\usepackage{bm}

\begin{document}

\title{An efficient implementation of two-component relativistic exact-decoupling methods for large molecules}

\author{Daoling Peng$^a$}

\author{Nils Middendorf$^b$}

\author{Florian Weigend$^{b,c}$}

\author{Markus Reiher$^a$}

\affiliation{$^a$ETH Zurich, Laboratorium f\"{u}r Physikalische Chemie,
Wolfgang-Pauli-Str.\ 10, CH-8093 Zurich, Switzerland}

\affiliation{$^b$Institute of Physical Chemistry, Karlsruhe Institute of Technology, Kaiserstr.\ 12, 76131 Karlsruhe, Germany}

\affiliation{$^c$Institute of Nanotechnology, Karlsruhe Institute of Technology, Hermann-von-Helmholtz-Platz 1, 76344 Eggenstein-Leopoldshafen, Germany}

\begin{abstract}
We present an efficient algorithm for one- and two-component relativistic exact-decoupling calculations.
Spin-orbit coupling is thus taken into account for the evaluation of relativistically transformed (one-electron) Hamiltonian.
As the relativistic decoupling transformation has to be evaluated with primitive functions,
the construction of the relativistic one-electron Hamiltonian becomes the bottleneck of the whole calculation for large molecules.
For the established exact-decoupling protocols, a minimal matrix operation count is established and discussed in detail.
Furthermore, we apply our recently developed local DLU scheme [J. Chem. Phys. 136 (2012) 244108] to accelerate this step.
With our new implementation two-component relativistic density functional calculations can be performed invoking the resolution-of-identity density-fitting approximation 
and (Abelian as well as non-Abelian) point group symmetry to accelerate 
both the exact-decoupling and the two-electron part.
The capability of our implementation is illustrated at the example of silver clusters with 
up to 309 atoms, for which the cohesive energy is calculated and extrapolated to the bulk.

\vspace{2cm}

{\bf Date:} April 18, 2013 \\
%{\bf Status:} revised version submitted to {\it J.\ Chem.\ Phys.}
{\bf Status:} published in {\it J. Chem. Phys.} 138, 184105 (2013); http://dx.doi.org/10.1063/1.4803693

\end{abstract}

\maketitle

\section{Introduction}\label{sec:intro}

Relativistic quantum chemistry is essential to the proper understanding
of the chemistry of any element in the periodic table with high
accuracy \cite{dyal07,pyyk08,reih09,bary10,wliu10_mp,saue11_cpc,auts12_jcp,mast10,wang12,pyyk12,schw13}.
Especially in heavy and super-heavy elements and their compounds,
relativistic effects largely determine to their electronic structures,
properties and functions.
The spin-orbit interaction is a relativistic effect, which
is very important for the calculation of spectroscopic constants (as, for example,
in electron spin resonance spectroscopy).
The relativistic four-component approach, in which the electronic Hamiltonian is constructed from one-electron Dirac operators
and two-electron Coulomb(--Breit) interaction operators, is able to provide very
accurate results for chemical problems.
However, it suffers from the presence of pathologic negative-energy solutions
and high computational cost.
To remove these drawbacks, a lower-cost relativistic electrons-only theory,
which provides a so-called two-component Hamiltonian from a unitary decoupling transformation,
is desirable for the description of molecular electronic structure.

Several relativistic two-component methods were developed in the past decades.
One of the widely used approaches is the second-order
Douglas--Kroll--Hess method (DKH2) \cite{hess86_pra,hess89_pra}.
It employs the free-particle Foldy--Wouthuysen (FW) \cite{fowo50_pr} transformation
as well as sequential Douglas--Kroll \cite{dokr74_ap} transformations
to decouple the four-component Hamiltonian.
Higher-order \cite{naka00_cpl,naka00_jcp,wolf02_jcp,vanw04_jcp}
and even arbitrary-order \cite{reih04_jcp1,reih04_jcp2,wolf06_jcp1,wolf06_jcp2,reih07b,peng09_jcp}
DKH methods have been developed.
The zeroth-order regular approximation (ZORA) \cite{chan86,vanl93_jcp,vanl94_jcp}
is another highly successful relativistic two-component method.
Within the ZORA framework, it is particularly easy to implement
the calculation of molecular properties;
see Refs.~\cite{auts11b,aqui11} for very recent examples and
Ref.~\cite{auts07} for a review.

The Barysz--Sadlej--Snijders (BSS) \cite{bary97_ijqc,bary01_jms,bary02_jcp,bary07_cpl}
method aims at exact decoupling of the free-particle-FW-transformed
four-component Hamiltonian by a unitary operator of the form derived in Ref.~\cite{heul86_jpb}.
This method has also been called IODK
(infinite-order Douglas--Kroll) \cite{sein08_cpl,sein10_jcp1,sein10_jcp2}.
Its Hamiltonian matrix is usually obtained by solving an iterative equation.
However, invoking the free-particle FW transformation turns out to be
not necessary for the construction of the exact-decoupling transformation
which has led to the formulation of a one-step protocol.
The pioneering work of this one-step transformation was provided by
Dyall \cite{dyal97_jcp,dyal98_jcp,dyal99_jcp,dyal01_jcp,dyal02_jcc}
in the form of the so-called normalized elimination of the small component (NESC) approach.
Later it was generalized to the so-called X2C method
by several groups \cite{fila03_jcp,jens05,fila05_jcp,
kutz05_jcp,kutz06_mp,wliu06_jcp,fila07_tca,wliu07_jcp,saue07_jcp,peng07_jcp,wliu09_jcp,sikk09_jcp}.
In contrast to the BSS method, the X2C Hamiltonian matrix is constructed non-iteratively.
The DKH method is also able to exactly decouple the four-component Hamiltonian matrix
and this has been shown within the arbitrary-order approach
\cite{reih04_jcp1,reih04_jcp2,wolf06_jcp1,wolf06_jcp2,mast07_jcp,peng09_jcp}.
For reviews of developments see
Refs.~\cite{hess00,wolf04,reih06_tca,wliu10_mp,naka11_cr,reih11_wire,saue11_cpc}

The formal and numerical comparison of exact decoupling approaches
was discussed in Ref.~\cite{peng12_tca}
which focused on theoretical aspects.
Recalling that the transformation should be performed in the uncontracted
basis representation, the cost of the relativistic part
is much higher than that of ordinary non-relativistic one-electron integral calculations
for which contracted basis functions can be used.
As we will show in this article, the relativistic part becomes
the bottleneck of fast density functional theory (DFT) calculations
especially in the two-component case
(without invoking the scalar-relativistic approximation).
We focus on the efficient implementation of relativistic
two-component approaches with acceleration techniques in this article.
The implementation details of exact-decoupling approaches
have been spread over many different papers.
We present here the necessary details with uniform notation
to describe our implementation.

Several techniques can be employed to improve the efficiency of constructing
the relativistic decoupling transformation.
The scalar structure of two-component operator matrices are presented
in detail in such a way that some matrix manipulations can be carried out at the
scalar level to avoid them at the general two-component level.
The basis representation is chosen such that some matrices are diagonal
as multiplication with diagonal matrices is computational very efficient.
The decoupling transformation matrices are evaluated in the original basis
to simplify the further reuse of them.
Symmetries of matrices are taken into account so that
only the symmetrically unique entries require explicit construction.
Other considerations such as point group symmetries and local approximations
are discussed as well.

The organization of this article is as follows.
Section \ref{sec:formalism} provides the formulas of exact-decoupling approaches
with minimum computational effort as a constraint.
In Section \ref{sec:formalism}, the implementation details and acceleration
schemes based on technical, symmetry, and physical considerations are discussed.
Numerical comparison of computation times and selected applications
are presented in Section \ref{sec:results}.
Finally, concluding remarks are described in Sec.~\ref{sec:conclusion}.

\section{Formalism}\label{sec:formalism}

We first discuss the equations of relativistic exact-decoupling approaches focusing on
how to minimize the computational requirements for our implementation.
The notation for bases and matrices is as follows:
Symbols formatted as $\mathsf{M}$ indicate real matrices in the basis of a set
of spin--free basis functions $\{\lambda_i\}$.
Formatting as $\bm{M}$ denotes matrices in
two-component (2c) spinor space $\{\chi_i\}$.
For matrix representations of four-component (4c) operators we use split notation for
large ($L$) and small ($S$) components explicitly if convenient.
To keep the notation in some places more compact, a notation like $\mathbb{M}$ is
used to indicate a $4\times 4$ super-matrix representation of a 4c operator.

The two-component electrons-only Hamiltonian is obtained from block-diagonalizing
the four-component (one-electron) Dirac equation (all expressions are given in atomic units; with the energy measured in units of Hartree, E$_{\text{H}}$)
\begin{equation}\label{moddiraceq}
\left(\begin{array}{cc}\bm{V}& \bm{T}\\ \bm{T}&(\frac{1}{4c^2}\bm{W}-\bm{T})\end{array}\right)
\left(\begin{array}{cc}\bm{C}_{L}^{+}&\bm{C}_{L}^{-}\\ \bm{C}_{S}^{+}&\bm{C}_{S}^{-}\end{array}\right)=
\left(\begin{array}{cc}\bm{S}&0\\ 0&\frac{1}{2c^2}\bm{T}\end{array}\right)
\left(\begin{array}{cc}\bm{C}_{L}^{+}&\bm{C}_{L}^{-}\\ \bm{C}_{S}^{+}&\bm{C}_{S}^{-}\end{array}\right)
\left(\begin{array}{cc}\bm{\epsilon}^{+}&0\\0&\bm{\epsilon}^{-}\end{array}\right)
\end{equation}
$\bm{V}$ denotes the
matrix representation of one-electron
potential-energy operator ($\mathcal{V}$) over two-component spinor functions,
$\bm{T}$ the non-relativistic kinetic energy matrix,
$\bm{S}$ the overlap matrix, and $\bm{W}$ the special relativistic potential matrix
\begin{equation}\label{defpvpmatrix}
\bm{W}_{ij}=\langle\chi_i|
\vec{\sigma}\cdot\vec{p}
\mathcal{V}
\vec{\sigma}\cdot\vec{p}
|\chi_j\rangle
\end{equation}
where $\vec{\sigma}$ denotes the vector of Pauli spin matrices,
$\vec{p}$ the linear momentum vector operator,
and $c$ the speed of light.
Eq.~(\ref{moddiraceq}) is the so-called modified Dirac equation \cite{dyal94_jcp},
which is the matrix Dirac equation employing the
kinetic balance (KB) condition \cite{ysle82_jcp,stat84_jcp,kutz84_ijqc,dyal84_jpb,dyal94_jcp,kutz97_cp} 
for the small components' basis set.
The KB condition ensures that one can obtain variationally stable
results.
The two-component electrons-only equation that is obtained from the
parent Eq.\ (\ref{moddiraceq}) reads
\begin{equation}\label{2ceoeq}
\bm{H}^{+}\bm{C}^{+}=\bm{S}\bm{C}^{+}\bm{\epsilon}^{+}.
\end{equation}
Note that the same metric $\bm{S}$ as in the
non-relativistic Schr\"{o}dinger-type equation is chosen.
The eigenvalues of Eq.~(\ref{2ceoeq}) are identical to the positive energy eigenvalues of
Eq.~(\ref{moddiraceq}) if exact decoupling has been achieved.
The positive energy eigenvalues are identified by verifying $\epsilon^{+}_{i}+c^2>0$
where $c^2$ represents the rest energy of an electron (in atomic units).

The two-component spinor functions $\{\chi_i\}$ are chosen as spin orbitals which
are the direct product of real scalar functions with spin functions $\{\lambda_i\}\otimes\{\alpha,\beta\}$.
Such a choice reduces the cost of basis orthogonalization since only real
matrices are involved and different spins are decoupled.
I.e., only the standard real matrices $\mathsf{S}$, $\mathsf{T}$, and $\mathsf{V}$
are needed 
\begin{eqnarray}\label{defstv}
\bm{S}=\left(\begin{array}{cc}\mathsf{S} & 0 \\ 0 & \mathsf{S}\end{array}\right),~\
\bm{T}=\left(\begin{array}{cc}\mathsf{T} & 0 \\ 0 & \mathsf{T}\end{array}\right),~\
\bm{V}=\left(\begin{array}{cc}\mathsf{V} & 0 \\ 0 & \mathsf{V}\end{array}\right).
\end{eqnarray}
$\bm{W}$ is a general complex matrix with the off-diagonal terms being non-zero.
But it can be evaluated from four real
matrices $\mathsf{W}^0,\mathsf{W}^x,\mathsf{W}^y,\mathsf{W}^z$,
\begin{equation}\label{defw}
\bm{W}=\left(\begin{array}{rr} \mathsf{W}^{0} + {\rm i} \mathsf{W}^z &~~ \mathsf{W}^y+ {\rm i} \mathsf{W}^x \\
-\mathsf{W}^y+ {\rm i} \mathsf{W}^x &~~ \mathsf{W}^{0} - {\rm i} \mathsf{W}^z \end{array}\right),
\end{equation}
where the individual real matrix elements read
\begin{subequations}
\begin{eqnarray}
\mathsf{W}^{0}_{ij}&=&\langle\lambda_i|
 p_x \mathcal{V}p_x
+p_y \mathcal{V}p_y
+p_z \mathcal{V}p_z |\lambda_j\rangle, \\
\mathsf{W}^{x}_{ij}&=&\langle\lambda_i|
 p_y \mathcal{V}p_z
-p_z \mathcal{V}p_y |\lambda_j\rangle, \\
\mathsf{W}^{y}_{ij}&=&\langle\lambda_i|
 p_z \mathcal{V}p_x
-p_x \mathcal{V}p_z |\lambda_j\rangle, \\
\mathsf{W}^{z}_{ij}&=&\langle\lambda_i|
 p_x \mathcal{V}p_y
-p_y \mathcal{V}p_x |\lambda_j\rangle.
\end{eqnarray}
\end{subequations}
$\mathsf{W}^{0}$ is a symmetric matrix and $\mathsf{W}^{x,y,z}$ are anti-symmetric matrices.
If the scalar-relativistic approximation is employed, one must neglect all terms
that result from spin-orbit (SO) coupling (i.e., $\mathsf{W}^{x,y,z}$).
Hence, the real
matrices ($\mathsf{S},\mathsf{T},\mathsf{V},\mathsf{W}^{0}$) instead of complex 2c matrices
can be employed to evaluate the scalar-relativistic Hamiltonian matrix.

The two-component, electrons-only, one-electron Hamiltonian matrix is a matrix function of the above-mentioned matrices,
\begin{equation}\label{r2ctrans}
\bm{H}=\bm{H}(\bm{S},\bm{T},\bm{V},\bm{W}).
\end{equation}
and no additional matrices of integrals are required for its evaluation apart from those just mentioned (we removed the superscript '$+$'
from here on since we do not discuss the Hamiltonian for negative energy solutions).
The decoupling transformation matrices $\bm{U}^{L}$ and $\bm{U}^{S}$ 
can also be obtained during the Hamiltonian construction procedure.
They are necessary for a later picture-change transformation of property integrals
as well as for the construction of local decoupling transformations.
The explicit decoupling transformation step to yield $\bm{H}$ reads
\begin{equation}\label{utrans}
\bm{H}=
\left(\begin{array}{cc} \bm{U}^{L,\dag} & \bm{U}^{S,\dag} \end{array}\right)
\left(\begin{array}{cc} \bm{V}&\bm{T} \\ \bm{T}&(\frac{1}{4c^2}\bm{W}-\bm{T})\end{array}\right)
\left(\begin{array}{c} \bm{U}^{L} \\ \bm{U}^{S} \end{array}\right).
\end{equation}
As the decoupling transformation yields not only the Hamiltonian matrix
but also the metric matrix in Eq.~(\ref{2ceoeq}), the following equation
\begin{equation}\label{renorm}
\bm{S}=
\left(\begin{array}{cc} \bm{U}^{L,\dag} & \bm{U}^{S,\dag} \end{array}\right)
\left(\begin{array}{cc} \bm{S} & 0 \\ 0 &  \frac{1}{2c^2}\bm{T} \end{array}\right)
\left(\begin{array}{c} \bm{U}^{L} \\ \bm{U}^{S} \end{array}\right)
\end{equation}
is satisfied as well.
This is the so-called renormalization condition which ensures that the relativistic
Hamiltonian matrix is diagonalized with the same metric as in the non-relativistic case.

Note that the basis contraction should be performed after the
relativistic matrix $\bm{H}$ was evaluated,
because the contraction of 4c spinors to efficiently describe the electronic orbitals
requires very different contraction coefficients for large and small components.
The same contraction scheme for both large and small components' bases (obeying the KB condition) 
cannot provide a reliable description for 4c spinors.
Thus, all input matrices ($\bm{S},\bm{T},\bm{V},\bm{W}$) should be calculated
using primitive functions instead of contracted functions.

\subsection{The X2C approach}

The X2C method employs an $X$-operator--based formula
\begin{equation}\label{u_x2c}
U_{\rm X2C}=
\left( \begin{array}{rr} \frac{\displaystyle 1}{\displaystyle \sqrt{1+X^{\dag}X}} &~
-X^{\dag}\frac{\displaystyle 1}{\displaystyle \sqrt{1+XX^{\dag}}} \\
X\frac{\displaystyle 1}{\displaystyle \sqrt{1+X^{\dag}X}} &~
\frac{\displaystyle 1}{\displaystyle \sqrt{1+XX^{\dag}}}\end{array}\right),
\end{equation}
for the exact decoupling unitary transformation.
Replacing the $X$ operator by its matrix representation in Eq.~(\ref{u_x2c})
yields the `matrix representation' of the X2C decoupling transformation.
However, this is valid only if orthonormal basis functions are employed.
The atom-centered Slater-type orbitals or Gaussian functions commonly used as bases for molecular
calculations do not form an orthogonal set.
Therefore, a matrix X2C decoupling transformation in a non-orthogonal
basis was proposed \cite{wliu09_jcp}.
With this non-orthogonal scheme, one must diagonalize the four-component Dirac equation
using a non-unit metric, i.e. solving Eq.~(\ref{moddiraceq}), which requires
a solver for a generalized eigenvalue problem.
However, the faster standard eigenvalue solver (with unit metric) can be employed
if an orthonormal representation is adopted.
Since the overlap matrix $\bm{S}$ and kinetic-energy-operator matrix $\bm{T}$
is composed of two identical real matrices
as shown in Eq.~(\ref{defstv}), one can diagonalize the real matrices
$\mathsf{S}$ and $\mathsf{T}$
instead of the larger metric matrix of Eq.~(\ref{moddiraceq})
to obtain orthonormalized basis functions.

In our implementation, we diagonalize the kinetic-energy-operator matrix $\mathsf{T}$ with metric $\mathsf{S}$
using a real symmetric generalized eigenvalue solver
\begin{equation}\label{orthonormalk}
\mathsf{T}\mathsf{K}=\mathsf{S}\mathsf{K}\mathsf{t},
\end{equation}
where $\mathsf{t}$ denotes the diagonal matrix of eigenvalues.
Note that this step is also the first one in a standard
DKH procedure \cite{wolf02_jcp}.
The basis transformation of the $\bm{T}$ matrix in the four-component equation is
simplified since $\bm{T}$ is diagonal in the $\mathsf{K}$-eigenfunction space.
The eigenvector matrix $\mathsf{K}$ has the following properties
\begin{eqnarray}\label{propk}
\mathsf{K}^{\dag}\mathsf{S}\mathsf{K}=\mathsf{I} ~~\mbox{and}~~
\mathsf{K}^{\dag}\mathsf{T}\mathsf{K}=\mathsf{t}.
\end{eqnarray}
In the following discussion,
it is convenient to introduce a diagonal matrix $\mathsf{p}$ (=$\sqrt{2\mathsf{t}}$)
which can be understood as the momentum eigenvalue matrix of the $\mathsf{K}$-eigenfunctions.
The 2c momentum-operator matrix $\bm{p}$ is obtained by duplicating the scalar
$\mathsf{p}$ to both spin components.

The non-unit metric of Eq.~(\ref{moddiraceq}) is then eliminated using the following
basis transformation matrix
\begin{eqnarray}\label{defk}
\mathbb{K}=\left(\begin{array}{cc}\bm{K}&0\\0&2c\bm{K}\bm{p}^{-1}\end{array}\right),
~~\mbox{where}~~
\bm{K}=\left(\begin{array}{cc}\mathsf{K}&0\\0&\mathsf{K}\end{array}\right).
\end{eqnarray}
However, the explicit construction of the matrix $\mathbb{K}$ is not required.
What we actually need is the 4c Hamiltonian matrix in the
orthonormal-basis representation
\begin{equation}
\mathbb{H}^{p}=\mathbb{K}^{\dag}\mathbb{H}\mathbb{K}=
\mathbb{K}^{\dag}\left(\begin{array}{cc}\bm{V}& \bm{T}\\
\bm{T}&(\frac{1}{4c^2}\bm{W}-\bm{T})\end{array}\right)\mathbb{K}=
\left(\begin{array}{cc}\widetilde{\bm{V}}&c\bm{p}\\c\bm{p}&
(\widetilde{\bm{W}}-2c^2)\end{array}\right),
\end{equation}
where
\begin{equation}
\widetilde{\bm{V}}=\bm{K}^{\dag}\bm{V}\bm{K},
\end{equation}
and
\begin{equation}
\widetilde{\bm{W}}=\bm{p}^{-1}\bm{K}^{\dag}\bm{W}\bm{K}\bm{p}^{-1}.
\end{equation}
Matrix multiplications to obtain $\widetilde{\bm{V}}$  are not performed
at the 2c level since $\widetilde{\bm{V}}$ has the same decoupled block-diagonal
structure as that in Eq.~(\ref{defstv}).
Only the scalar term $\mathsf{K}^{\dag}\mathsf{V}\mathsf{K}$
is needed to construct $\widetilde{\bm{V}}$.
For the special relativistic potential matrix $\widetilde{\bm{W}}$, four real matrices
\begin{equation}
\widetilde{\mathsf{W}}^{q}=\mathsf{p}^{-1}\mathsf{K}^{\dag}\mathsf{W}^{q}\mathsf{K}\mathsf{p}^{-1},~\
q\text{ in }\{0,x,y,z\},
\end{equation}
are required, and $\widetilde{\bm{W}}$ is obtained using the form in Eq.~(\ref{defw})
with $\mathsf{W}^{q}$ replaced by $\widetilde{\mathsf{W}}^{q}$.

Once the basis functions are converted to the orthonormal-basis representation, a standard
hermitian eigenvalue solver can be employed to diagonalize the 4c Hamiltonian matrix
\begin{equation}\label{diaghdp}
\mathbb{H}^{p}
\left(\begin{array}{c}\bm{C}_{L'}^{+} \\ \bm{C}_{S'}^{+} \end{array}\right)=
\left(\begin{array}{c}\bm{C}_{L'}^{+} \\ \bm{C}_{S'}^{+} \end{array}\right)
\bm{\epsilon}^{+},
\end{equation}
where we only consider the positive energy solutions.
Since a basis transformation was applied, the `large' ($L'$) and `small' ($S'$)
components are no longer the original large and small components of the modified
Dirac equation, but they are connected by the transformation matrix $\mathbb{K}$.
The matrix representation of the $X'$ operator, which converts the large component
to the small component in the orthonormal-basis representation, is evaluated from
the coefficients of positive energy solutions
\begin{equation}\label{defxortho}
\bm{X}'=\bm{C}_{S'}^{+}\left(\bm{C}_{L'}^{+}\right)^{-1}.
\end{equation}
The renormalization matrix $\bm{R}'$ is obtained as follows
\begin{equation}\label{defrenorm}
\bm{R}'=(\bm{I}+\bm{X}'^{\dag}\bm{X}')^{-\frac{1}{2}},
\end{equation}
where $1/2$ denotes the principle square root of a positive-definite matrix.
`Principle' means that the positive roots instead of negative roots were chosen to
compute the square root of a matrix.
It is simple to verify that $\bm{I}+\bm{X}'^{\dag}\bm{X}'$ is positive-definite.
$\bm{R}'$ is therefore uniquely defined.
The final relativistic electrons-only Hamiltonian matrix then reads
\begin{equation}
\bm{H}=(\bm{R}'\bm{K}^{-1})^{\dag}\left(
\widetilde{\bm{V}}+c\bm{p}\bm{X}'+c\bm{X}'^{\dag}\bm{p}+
\bm{X}'^{\dag}(\widetilde{\bm{W}}-2c^2)\bm{X}'
\right)\bm{R}'\bm{K}^{-1},
\end{equation}
where $\bm{K}^{-1}$ is the back transformation to original basis representation.

To calculate the picture-change correction of property operators or employ
the local relativistic approximation \cite{peng12_jcp},
one needs to store the decoupling transformation matrices in the
original basis representation for further use.
The X2C decoupling transformation matrices are calculated as follows
\begin{subequations}\label{ulsortho}
\begin{eqnarray}
\bm{U}_{\rm X2C}^{L}&=&\bm{K}\bm{R}'\bm{K}^{-1},\label{ulortho}\\
\bm{U}_{\rm X2C}^{S}&=&2c\bm{K}\bm{p}^{-1}\bm{X}'\bm{R}'\bm{K}^{-1}.\label{usortho}
\end{eqnarray}
\end{subequations}
With Eqs.~(\ref{propk}) and (\ref{defrenorm}) one
can verify that the renormalization condition Eq.~(\ref{renorm}) is satisfied
with the X2C decoupling transformation matrices.
The above formula seems very different compared to the expression in our previous
implementations \cite{wliu09_jcp,peng12_tca,peng12_jctc}
\begin{subequations}\label{ulsnonortho}
\begin{eqnarray}
\bm{U}_{\rm X2C}^{L}&=&\bm{S}^{-\frac{1}{2}}\left(\bm{S}^{-\frac{1}{2}}
(\bm{S}+\frac{1}{2c^2}\bm{X}^{\dag}\bm{T}\bm{X})
\bm{S}^{-\frac{1}{2}}\right)^{-\frac{1}{2}}\bm{S}^{\frac{1}{2}},\label{ulnonortho}\\
\bm{U}_{\rm X2C}^{S}&=&\bm{X}\bm{U}_{\rm X2C}^{L},\label{usnonortho}
\end{eqnarray}
\end{subequations}
which employs the $\bm{X}$ matrix defined in the original non-orthogonal
basis representation.
However, these two forms can be shown to be equivalent
(see Appendix \ref{secapp}).
We adopt the form of Eq.~(\ref{ulsortho}) which employs the $\bm{X}'$ matrix in 
orthonormal-basis representation, because it is more efficient and requires fewer matrix manipulations.

\subsection{The BSS approach}

In the BSS approach, the free-particle FW transformation
in addition to the orthonormal transformation $\mathbb{K}$
is applied to obtain the four-component Hamiltonian matrix to be diagonalized.
The free-particle FW transformation $\mathbb{U}_0$ features four diagonal block matrices,
\begin{equation}
\mathbb{U}_0=
\left( \begin{array}{rr}
\bm{U}^{A}_{0} &~ \bm{U}^{B}_{0} \\
-\bm{U}^{B}_{0} & \bm{U}^{A}_{0} \end{array}\right)
=\left(\begin{array}{cc} \sqrt{\frac{\displaystyle \bm{E}_0+c^2}{\displaystyle 2\bm{E}_0}}  &
\sqrt{\frac{\displaystyle \bm{E}_0-c^2}{\displaystyle 2\bm{E}_0}} \\ -\sqrt{\frac{\displaystyle \bm{E}_0-c^2}{\displaystyle 2\bm{E}_0}} &
\sqrt{\frac{\displaystyle \bm{E}_0+c^2}{\displaystyle 2\bm{E}_0}} \end{array}\right),
\end{equation}
with $\bm{E}_0=\sqrt{c^2\bm{p}^2+c^4}$.
It is then applied to yield a transformed four-component Hamiltonian
matrix $\mathbb{H}^{0}$
\begin{equation}
\mathbb{H}^{0}=\mathbb{U}_0^{\dag}\mathbb{K}^{\dag}
\mathbb{H}\mathbb{K}\mathbb{U}_0
=\left(\begin{array}{ll}
\bm{U}^{A}_{0}\widetilde{\bm{V}}\bm{U}^{A}_{0}+\bm{U}^{B}_{0}\widetilde{\bm{W}}\bm{U}^{B}_{0}
+\bm{E}_{0}-c^2 & ~~
\bm{U}^{A}_{0}\widetilde{\bm{V}}\bm{U}^{B}_{0}-\bm{U}^{B}_{0}\widetilde{\bm{W}}\bm{U}^{A}_{0} \\
\bm{U}^{B}_{0}\widetilde{\bm{V}}\bm{U}^{A}_{0}-\bm{U}^{A}_{0}\widetilde{\bm{W}}\bm{U}^{B}_{0} & ~~
\bm{U}^{B}_{0}\widetilde{\bm{V}}\bm{U}^{B}_{0}+\bm{U}^{A}_{0}\widetilde{\bm{W}}\bm{U}^{A}_{0}
-\bm{E}_{0}-c^2\end{array}\right)
.
\end{equation}
Since $\mathbb{U}_0$ is a unitary matrix, it preserves
the orthonormality condition. $\mathbb{H}^{0}$ is then diagonalized by a standard
hermitian eigenvalue solver.
The eigenvalue equation has the same structure as Eq.~(\ref{diaghdp}) with
labels $L'$ and $S'$ replaced by $L''$ and $S''$.
The $\bm{X}''$ matrix in this basis representation is obtained by
Eq.~(\ref{defxortho}) with same label replacement.

Next, the exact-decoupling BSS transformation
\begin{equation}
\label{u1trafo}
\mathbb{U}_{1}=
\left( \begin{array}{rr} \frac{\displaystyle 1}{\displaystyle \sqrt{\bm{I}+\bm{X}''^{\dag}\bm{X}''}} &~
-\bm{X}''^{\dag}\frac{\displaystyle 1}{\displaystyle \sqrt{\bm{I}+\bm{X}''\bm{X}''^{\dag}}} \\
\bm{X}''\frac{\displaystyle 1}{\displaystyle \sqrt{\bm{I}+\bm{X}''^{\dag}\bm{X}''}} &
\frac{\displaystyle 1}{\displaystyle \sqrt{\bm{I}+\bm{X}''\bm{X}''^{\dag}}}\end{array}\right),
\end{equation}
which has the same structure as the exact-decoupling transformation $U_{\rm X2C}$,
is applied.
After the exact decoupling BSS transformation has been carried out, the Hamiltonian
matrix is back-transformed to the original non-orthogonal basis representation
\begin{equation}
\mathbb{H}=(\mathbb{K}^{-1})^{\dag}
\mathbb{U}_{1}^{\dag}\mathbb{H}^{0}\mathbb{U}_{1}\mathbb{K}^{-1}.
\end{equation}
However, only the electronic (upper-left) part of the 4c matrix $\mathbb{H}$
needs to be evaluated. In applying the above transformation,
we discard the second column of $\mathbb{U}_{1}$ in Eq.\ (\ref{u1trafo})
and replace $\mathbb{K}$ by $\bm{K}$.
The BSS decoupling transformation matrices
$\bm{U}_{\rm BSS}^{L}$ and $\bm{U}_{\rm BSS}^{S}$ read
\begin{subequations}
\begin{eqnarray}
\bm{U}_{\rm BSS}^{L}&=&\bm{K}(\bm{U}^{B}_{0}\bm{X}''+\bm{U}^{A}_{0})
\bm{R}''\bm{K}^{-1},\\
\bm{U}_{\rm BSS}^{S}&=&2c\bm{K}\bm{p}^{-1}
(\bm{U}^{A}_{0}\bm{X}''-\bm{U}^{B}_{0})
\bm{R}''\bm{K}^{-1},
\end{eqnarray}
\end{subequations}
where $\bm{R}''$ denotes $(\bm{I}+\bm{X}''^{\dag}\bm{X}'')^{-\frac{1}{2}}$.

Comparing to the X2C approach, the BSS approach requires a few more matrix multiplications.
The off-diagonal terms of the X2C Hamiltonian matrix $\mathbb{H}^{p}$ are
diagonal matrices, while the ones in the BSS Hamiltonian matrix $\mathbb{H}^{0}$ are not.
It then requires more matrix multiplications
to take into account the non-diagonal feature.
Furthermore, the BSS decoupling transformation matrices are composed of
more terms than the X2C expression in Eq.~(\ref{ulsortho}).
This, of course, introduces more matrix multiplications in an implementation.

\subsection{The DKH approach}

The DKH approach requires the free-particle FW transformation as the initial transformation \cite{reih04_jcp1}
to obtain the transformed Hamiltonian matrix $\mathbb{H}^{0}$,
which is composed of even terms (i.e., diagonal terms, denoted as $\bm{E}$) and
odd terms (i.e., off-diagonal terms, denoted as $\bm{O}$)
with the subscripts denoting the order in the external potential $\mathcal{V}$
\begin{equation}
\mathbb{H}^{0}=\left(\begin{array}{cc}\bm{E}_{0}-c^2&0\\0&-\bm{E}_{0}-c^2\end{array}\right)
+\left(\begin{array}{cc}\bm{E}_{1}&\bm{O}_{1}\\ \bm{O}^{\dag}_{1}&\bm{E}'_{1}\end{array}\right).
\end{equation}
Subsequent decoupling transformations are expressed as
\begin{equation}
\mathbb{U}^{(n)}=\prod_{k=1}^{n}\mathbb{U}_{k}
\end{equation}
with the generalized parametrization of the $\mathbb{U}_k$ \cite{wolf02_jcp},
\begin{equation}
\mathbb{U}_k=\sum_{i=0}^{[n/k]}a_{k,i}\mathbb{W}_{k}^{i}
=\sum_{i=0}^{[n/k]}a_{k,i}
\left(\begin{array}{cc}0&\bm{W}_{k}\\ -\bm{W}^{\dag}_{k}&0\end{array}\right)^{i},
\end{equation}
in terms of anti-hermitian matrix operators $\mathbb{W}_{k}$.
Here, $n$ is the order of the DKH expansion.
To keep the convention for the anti-hermitian matrix consistent with the literature,
the same letter '$W$' is used in this subsection, while
noting that it is not the relativistic potential-energy integral defined in Eq.~(\ref{defw}).
The polynomial cost algorithm \cite{peng09_jcp} for the evaluation of the
anti-hermitian matrices $\mathbb{W}_{k}$ and for the DKH Hamiltonian was
employed.
For the different parametrization schemes of the $\mathbb{U}_k$ \cite{wolf02_jcp},
the exponential parametrization was chosen since it requires the lowest number
of matrix multiplications \cite{peng09_jcp}.

However, the number of matrix multiplications can be further reduced  by
two considerations.
On the one hand, the intermediate operator products which do not contribute to the
final DKH Hamiltonian can be neglected.
For example, in the $k$-th step the $\mathbb{W}_{k}$ matrix is multiplied to
an intermediate $\mathbb{M}_{l}$ of order $l$ in the external potential.
If $k+2l>n\geq k+l$ and $\mathbb{M}_{l}$ is even, the multiplication with $\mathbb{W}_{k}$
can be skipped. Because the intermediate term, which is the product of
$\mathbb{W}_{k}$ and $\mathbb{M}_{l}$, is odd and then does not contribute to
the $n$-th order DKH Hamiltonian.
The further multiplication to $\mathbb{W}_{k}$ yields an even matrix
but goes beyond $n$-th order.
Furthermore, the DKH Hamiltonian matrix is taken from the upper part
of the 4c matrix while the lower part
(which is not identical to the upper part in our algorithm)
is not required.
For instance, if $k+l=n$ and $\mathbb{M}_{l}$ is odd, the product with
$\mathbb{W}_{k}$, of course, contributes to the final DKH Hamiltonian but the
matrix multiplications to obtain the lower part result can be neglected.

On the other hand, the symmetry of the matrices can be exploited.
Noting that the odd matrices $\mathbb{O}$ are hermitian
and $\mathbb{W}$ are anti-hermitian.
The algorithm of exponential parametrization requires the evaluation of their
commutator
\begin{equation}
\left[\left(\begin{array}{cc}0&\bm{W}\\ -\bm{W}^{\dag}&0\end{array}\right),
\left(\begin{array}{cc}0&\bm{O}\\ \bm{O}^{\dag}&0\end{array}\right)\right]=
\left(\begin{array}{cc} \bm{W}\bm{O}^{\dag}+(\bm{W}\bm{O}^{\dag})^{\dag}&0\\
0&-\bm{W}^{\dag}\bm{O}-(\bm{W}^{\dag}\bm{O})^{\dag}\end{array}\right).
\end{equation}
According to the above equation, two instead of four matrix
multiplications of 2c matrices are enough to evaluate the commutator.
For example, the upper part is the sum of $\bm{W}\bm{O}^{\dag}$ and its
hermitian transpose which does not require the computation of its
matrix product form $\bm{O}\bm{W}^{\dag}$.
The commutator of $\mathbb{W}$ with even matrices $\mathbb{E}$ reads
\begin{equation}
\left[\left(\begin{array}{cc}0&\bm{W}\\ -\bm{W}^{\dag}&0\end{array}\right),
\left(\begin{array}{cc}\bm{E}&0\\0&\bm{E}'\end{array}\right)\right]=
\left(\begin{array}{cc}0&\bm{W}\bm{E}'-\bm{E}\bm{W}\\
(\bm{W}\bm{E}'-\bm{E}\bm{W})^{\dag}&0\end{array}\right).
\end{equation}
The lower part is just the conjugate transpose of the upper part and
thus does not need explicit construction.
Therefore, only $\{\bm{O},\bm{W},\bm{E},\bm{E}'\}$ need to be
calculated and stored in the DKH-Hamiltonian evaluation procedure.

With these considerations for the reduction of the computational cost,
the number of 2c matrix multiplications required for the construction of
the DKH Hamiltonian of orders from 2 to 14 are listed in Table~\ref{tab_nmmdkh}.
The corresponding data without these considerations is taken from
Ref.~\cite{peng09_jcp}.
It is evident from Table~\ref{tab_nmmdkh} that
the number of matrix multiplications is significantly decreased.
The number of multiplications for lower-order DKH is surprisingly small.
For example, DKH2 requires only one and DKH3 requires three more.
The explicit simplified formulas for DKH2 and DKH3 are
\begin{eqnarray}
\bm{E}_2&=&\frac{1}{2}(\bm{W}_1\bm{O}_1^{\dag}+c.t.),\\
\bm{E}_3&=&\frac{1}{2}\left(\bm{W}_1(\bm{W}_1\bm{E}_1
-\bm{E}'_1\bm{W}_1)^{\dag}+c.t.\right),
\end{eqnarray}
where $c.t.$ denotes the conjugate transpose of the former term.
Since $\bm{W}_1$ is evaluated from $\bm{O}_1$ multiplying with kinematic factors,
it does not require a matrix multiplication.
It is then clear that only one matrix multiplication is necessary for DKH2
and four are necessary for DKH3
(noting $\bm{H}_{\rm DKH3}=\sum_{i=0}^{3}\bm{E}_{i}-c^2$).

%---------------------------------------------------------------------------------------------
\begin{table}[H]
\caption{Number of matrix multiplications $P(n)$ required
 for the evaluation of DKH$n$ Hamiltonians with an exponential parametrization
of the unitary transformation. The right column shows the actual operation count established in this work. \label{tab_nmmdkh}}
\begin{center}
\begin{tabular}{rrr}
\hline\hline
$n$ & Ref.~\cite{peng09_jcp}& this work \\
\hline
 2 &    8 &   1 \\
 3 &   16 &   4 \\
 4 &   36 &   9 \\
 5 &   56 &  17 \\
 6 &   96 &  26 \\
 7 &  136 &  38 \\
 8 &  200 &  55 \\
 9 &  264 &  79 \\
10 &  360 & 104 \\
11 &  448 & 132 \\
12 &  576 & 169 \\
13 &  700 & 217 \\
14 &  860 & 266 \\
\hline\hline
\end{tabular}
\end{center}
\end{table}
%---------------------------------------------------------------------------------------------

If $n$ is large (strictly, if it approaches infinity), exact decoupling is achieved.
Usually, a very low value for $n$ is sufficient for calculations of
relative energies and valence-shell properties.
With the DKH decoupling transformation written as
\begin{equation}
\mathbb{U}^{(n)}=\left(\begin{array}{cc}\bm{U}^{(n),LL}&\bm{U}^{(n),LS}\\
\bm{U}^{(n),SL}&\bm{U}^{(n),SS}\end{array}\right),
\end{equation}
$\bm{U}_{\rm DKH}^{L}$ and $\bm{U}_{\rm DKH}^{S}$ read
\begin{subequations}\label{dkhtrans}
\begin{eqnarray}
\bm{U}_{\rm DKH}^{L}&=&\bm{K}(\bm{U}^{B}_{0}\bm{U}^{(n),SL}+
\bm{U}^{A}_{0}\bm{U}^{(n),LL})\bm{R}''\bm{K}^{-1},\\
\bm{U}_{\rm DKH}^{S}&=&2c\bm{K}\bm{p}^{-1}(\bm{U}^{A}_{0}\bm{U}^{(n),SL}-
\bm{U}^{B}_{0}\bm{U}^{(n),LL})\bm{R}''\bm{K}^{-1}.
\end{eqnarray}
\end{subequations}
Since only $LL$ and $SL$ components of $\mathbb{U}^{(n)}$ are required,
computations needed for the evaluation of other components can be neglected.

\section{Implementation}\label{sec:implementation}

The number of matrix operations necessary for the implementation of different
two-component approaches presented in
Section~\ref{sec:formalism} has been collected in Table~\ref{tab_matrixops} (for comparison,
we provide the operation count for the {\sc scalar-relativistic} variants in Table \ref{tab_matrixops_1c}).
Because the multiplication with diagonal matrices
requires much fewer multiplications of numbers
than the ordinary matrix multiplication,
it is not counted in Table~\ref{tab_matrixops}.
The multiplication of a general matrix with a diagonal matrix requires
$\mathcal{O}(M^2)$ multiplications of numbers,
where $M$ denotes the dimension of the matrix,
while the scaling of the multiplication of two general matrices
is formally $\mathcal{O}(M^3)$.
If $M$ is large, the cost of the former case is negligible.

%---------------------------------------------------------------------------------------------
\begin{table}[H]
\caption{Number of matrix operations necessary for the implementation
of {\it two-component} relativistic exact-decoupling approaches. $N$ denotes the number of scalar
basis functions. P($n$) is given in Table~\ref{tab_nmmdkh}.}
\label{tab_matrixops}
\begin{center}
\begin{tabular}{lclccc}
\hline\hline
Operation & Dimension & Type & X2C & BSS & DKH$n$ \\
\hline
Diagonalization &  $N$ & Real      &  1 &  1 & 1 \\
Diagonalization & $2N$ & Complex   &  1 &  1 & 0 \\
Diagonalization & $4N$ & Complex   &  1 &  1 & 0 \\
Inversion       &  $N$ & Real      &  1 &  1 & 1 \\
Inversion       & $2N$ & Complex   &  1 &  1 & 0 \\
Multiplication  &  $N$ & Real      & 10 & 10 & 10 \\
Multiplication  & $2N$ & Complex   & 11 & 14 & 2+P($n$)\\
\hline\hline
\end{tabular}
\end{center}
\end{table}
%---------------------------------------------------------------------------------------------

%---------------------------------------------------------------------------------------
\begin{table}[H]
\caption{Number of matrix operations necessary for the implementation
of {\it scalar} relativistic exact-decoupling approaches. $N$ denotes the
number of scalar basis functions. Note that all matrices are real in this case. P($n$) is given in Table~\ref{tab_nmmdkh}.}
\label{tab_matrixops_1c}
\begin{center}
\begin{tabular}{lcccc}
\hline\hline
Operation & Dimension & X2C & BSS & DKH$n$ \\
\hline
Diagonalization &  $N$ &  2 &  2 & 1 \\
Diagonalization & $2N$ &  1 &  1 & 0 \\
Inversion       &  $N$ &  2 &  2 & 1 \\
Multiplication  &  $N$ & 15 & 18 & 6+P($n$) \\
\hline\hline
\end{tabular}
\end{center}
\end{table}
%---------------------------------------------------------------------------------------

Besides matrix multiplication,
matrix diagonalization and inversion are required.
Both of them are also of $\mathcal{O}(M^3)$ cost.
As we can see from Table~\ref{tab_matrixops},
the X2C and BSS approach require the same five different types of matrix diagonalization
and inversion, while the DKH approach requires only two of them.
The commonly used two operations are matrix diagonalization of dimension $N$,
which is used for basis orthonormalization as in Eq.~(\ref{orthonormalk}),
and matrix inversion of dimension $N$ which is the calculation of
the inverse basis transformation matrix $\mathsf{K}^{-1}$.
$N$ denotes the number of scalar basis functions.

Both the X2C and BSS approach employ a 4c eigenvalue equation to obtain the
so-called $\bm{X}$ matrix, a $4N$-dimensional complex matrix diagonalization is
then needed for both of them.
The complex matrix inversion of dimension $2N$ is employed to calculate the
$\bm{X}$ matrix using Eq.~(\ref{defxortho}).
The remaining $2N$-dimensional complex matrix diagonalization computes the
inverse square root of a hermitian matrix and it is the main effort
in the calculation of
the renormalization matrix $\bm{R}$ according to Eq.~(\ref{defrenorm}).
The procedure is illustrated as follows.
Given the hermitian matrix $\bm{A}$,
its eigenvectors and eigenvalues are contained in
$\bm{C}$ and in the diagonal matrix $\bm{a}$, respectively,
\begin{equation}
\bm{A}\bm{C}=\bm{C}\bm{a}.
\end{equation}
The inverse (principle) square root of the matrix $\bm{A}$ is then computed by
\begin{equation}
\bm{A}^{-\frac{1}{2}}=\bm{C}\bm{a}^{-\frac{1}{2}}\bm{C}^{\dag}.
\end{equation}

The ten real matrix multiplications that are commonly required for
our implementation of all exact-decoupling approaches are
the orthonormal basis transformation of five potential-energy matrices
\begin{equation}
\mathsf{K}^{\dag}\mathsf{A}\mathsf{K},\ \text{ with }\mathsf{A}\text{ in }
\{\mathsf{V},\mathsf{W}^0,\mathsf{W}^x,\mathsf{W}^y,\mathsf{W}^z\}.
\end{equation}
The number of 2c complex matrix multiplications is different for different approaches.
The BSS approach differs from the X2C approach only in this term.
It requires three more than the X2C approach since the off-diagonal parts of its 4c
Hamiltonian matrix are not diagonal.
For the DKH approach, the number of 2c complex matrix multiplications
depends on the order of the expansion.
As we discussed in Section~\ref{sec:formalism},
it could be a very small number, for instance it is only three for DKH2.
Consequently, the low-order DKH method is much faster than the X2C method.
The numerical comparison of computation times for different approaches
will be presented in Section~\ref{sec:results}.

If the scalar approximation (neglecting spin-orbit coupling terms)
is employed, Table~\ref{tab_matrixops} would become quite different.
The operation count for the scalar-relativistic variants is given in Table \ref{tab_matrixops_1c}.
Firstly, all complex numbers become real since one could employ real basis functions
and the Hamiltonian operators are real as well.
Thus, only real matrix operations are required.
Secondly, spin symmetry can be used so that dimensions of all 2c and 4c matrices
are reduced by half.
Finally, since the spin-orbit components of the relativistic potential matrices
($\mathsf{W}^x,\mathsf{W}^y,\mathsf{W}^z$) are neglected, the number of matrix
multiplications required for the orthonormal basis transformation is then decreased
from ten to four.

\subsection{Acceleration schemes}

Apparently, the evaluation of the relativistic Hamiltonian matrix requires only few
matrix multiplications. Hence, it should be computationally less demanding than
the self-consistent field (SCF) iterations.
However, the basis functions must be used in uncontracted form during the
relativistic set-up. The dimension of relativistic matrices is then much
larger than that of Fock matrices in a contracted basis.
The calculation of the relativistic one-electron Hamiltonian can thus become the bottleneck of a whole calculation,
especially in the case of heavily contracted basis functions
and fast DFT techniques employed for SCF calculations.
Therefore, methods to accelerate
the calculation of the relativistic transformation must be taken into account.

Of course, parallelization can be employed to reduce the computation time using
multi-processor or multi-core hardware.
This could be simply achieved by integrating a parallel library for matrix algebra
since almost all cost of the relativistic transformations are carried out by matrix
manipulations.
However, it is efficient only if the dimension of matrices is large.

Many symmetries can be exploited for quantum chemical calculations, some of them
can also be applied to the relativistic transformation.
First of all, Hamiltonian matrices and some intermediate matrices are
hermitian or real symmetric.
Therefore, special matrix routines which take the matrix symmetry into account
can be invoked to reduce the computational cost.
For molecules having point group symmetries,
relativistic transformations are performed within each irreducible
representation. The matrix block of two different irreducible representations
is always zero and can thus be skipped.
We present the implementation details of point group symmetry in the next subsection.

Time reversal symmetry, which is a reminiscence of the double occupation of spatial orbitals in non-relativistic theory, can be exploited
for two-component calculations. In order to exploit this symmetry, Kramers pairs,
which are connected by the time reversal operator, must be employed as basis.
An operator in a Kramers-paired basis has the following structure
\begin{equation}
\left(\begin{array}{cc} A & B \\ -B^{*} & A^{*} \end{array}\right),
\end{equation}
where $A$ and $B$ are general complex matrices. 
One can only store
the upper part of the above time-reversal symmetric matrix and
the lower part can be generated on-the-fly when required.
Time reversal symmetry is fully compatible with the double
point group symmetry, i.e., one can simultaneously exploit both point group symmetry
and time reversal symmetry,
see Refs.~\cite{peng09_ijqc,peng11_tca} for details.
Time reversal symmetry reduces the computational cost by half.

Acceleration schemes discussed above are methods without any loss of accuracy.
Based on physical considerations, very small quantities can be neglected
to accelerate the calculation at the cost of a negligible loss of accuracy.
For example, an integral involving two distant atoms is usually very close to zero.
We can then introduce an approximation to neglect such small terms.
The relativistic effect is roughly proportional to
$Z^2$ where $Z$ is the atomic number of the elements under consideration.
Therefore, applying the relativistic transformation only for
heavy atoms of a molecule should be reasonable
and was studied in
the diagonal local approximation to the one-electron Hamiltonian (DLH) \cite{pera04_jcp,pera05_jcp,thar09_jcp,peng12_jcp}.
However, the DLH approximation completely neglects the relativistic corrections to
atom--other-atom terms and might not be accurate enough.
We have proposed the diagonal local approximation to the unitary transformation (DLU) \cite{peng12_jcp} instead, which
takes into account the relativistic treatment of the atom--other-atom terms
with excellent accuracy.
A similar local approximation was developed for the BSS transformation of
one-electron \cite{sein12_jcp1} and two-electron \cite{sein12_jcp2} operators.
By contrast, the DLU scheme was developed and applied to the X2C, BSS, and DKH approaches.

The idea of the DLU approximation is to approximate the decoupling transformation matrices
$\bm{U}^{L}$ and $\bm{U}^{S}$ as the direct sum of atomic blocks only,
\begin{subequations}\label{dlutrans}
\begin{eqnarray}
\bm{U}^{L}&=&\bigoplus_A \bm{U}^{L}_{AA}, \label{localUL}\\
\bm{U}^{S}&=&\bigoplus_A \bm{U}^{S}_{AA}, \label{localUS}
\end{eqnarray}
\end{subequations}
in such a way that a block-diagonal approximation to the unitary matrix is obtained
where $A$ labels the atom-centered basis functions of a specific atom $A$. The atomic components of the decoupling
transformation matrices are evaluated within each atomic block
of atom-centered basis functions
\begin{equation}\label{r2ctransatom}
\bm{U}^{L,S}_{AA}=\bm{U}^{L,S}(\bm{S}_{AA},\bm{T}_{AA},\bm{V}_{AA},\bm{W}_{AA}),
\end{equation}
employing the same relativistic matrix function
with different input of atomic matrices.
In contrast to the relativistic transformation at full molecular dimension,
only small pieces of the input matrices are required.
The cost of the relativistic transformation is decreased significantly
by the DLU approximation, especially if the number of atoms is large.
The matrix functions, i.e., the construction of the relativistic Hamiltonian matrix and
decoupling transformation matrices, are as discussed in Section~\ref{sec:formalism}.
Therefore, all relativistic decoupling approaches are compatible with the DLU approximation.
The final molecular Hamiltonian matrix is obtained by
applying the DLU decoupling transformation Eq.~(\ref{dlutrans}) using Eq.~(\ref{utrans}).
One can further reduce the computational cost by invoking the non-relativistic
approximation (setting atomic decoupling transformation matrices to identity matrices)
to all light atoms such as hydrogen \cite{peng12_jcp}.
The neighboring-atomic-blocks approximation can be used to achieve linear scaling \cite{peng12_jcp}.
It turns out that these approximations can be expected to give negligible errors \cite{peng12_jcp} (but see also the discussion below).

\subsection{Exploitation of point group symmetry}

Exploitation of molecular point group symmetry is one of the prominent
features in {\sc Turbomole} \cite{turbomole}.
Large computational savings are achieved in the SCF loop by, e.g., the restriction to
non-redundant grid points for the DFT part, non-redundant basis and auxiliary basis
functions in the Coulomb part, and further by the usage of symmetry adapted basis functions,
which allows for blocking according to irreducible representations.
On the other hand, in this way the relativistic transformation
procedure may easily become the time-dominating step for symmetric molecules,
unless symmetry is exploited also there, e.g., by employing symmetry
adapted combinations (SAOs) of the primitive basis functions (AOs).
This is organized as follows.

At first, the transformation AO-SAO coefficients $\mathbf{c}_{\text{AS}}$ for
primitive functions are obtained by routines already existing in {\sc Turbomole}, which construct symmetry-adapted 
shells from symmetry-non-redundant shells and their symmetry images by means of elementary group theory.
Next, matrices $\mathbf{S}$ and $\mathbf{W}$ are calculated in the AO basis
and transformed to the SAO basis with those coefficients.
In this way, a block-diagonal form is obtained
(matrix elements of basis functions belonging to different irreducible representations are zero),
and the relativistic transformation procedure can be carried out in the
SAO basis for each block separately.
This yields the relativistic one-electron contributions, $\mathbf{H}_{\text{SS}}$,
as a block-diagonal matrix in the basis of symmetry adapted primitive basis functions.
The transformation to the contracted basis used in the rest of the program has
to be done in the (not symmetry adapted) AO basis,
so $\mathbf{H}_{\text{SS}}$ has to be transformed to $\mathbf{H}_{\text{AA}}$.
For this, the inverse coefficients $\mathbf{c}^{-1}_{\text{AS}}$ are needed,
\begin{align}
\mathbf{c}^{-1}_{AS}=\mathbf{S}^{-1}_{\text{SS}}\mathbf{S}_{\text{SA}}.
\end{align}
For this purpose the overlap matrix calculated in the basis of primitive basis functions,
$\mathbf{S}_{\text{AA}}$, has to be transformed to the SAO basis
leading to a block-diagonal form, $\mathbf{S}_{\text{SS}}$.
The inversion is carried out for each block separately.
The resulting matrix, $\mathbf{S}^{-1}_{\text{SS}}$,
is multiplied by $\mathbf{S}_{\text{SA}}$, the (non-inverted) overlap matrix
with only one index transformed to the SAO basis.
It has to be noted that the resulting coefficient matrix $\mathbf{c}^{-1}_{\text{AS}}$
is by far not as sparse as $\mathbf{c}_{\text{AS}}$. Thus, the cheapest way for the transformation
from $\mathbf{H}_{\text{AA}}$ to $\mathbf{H}_{\text{SS}}$ are ordinary matrix multiplications.
In this transformation step symmetry is not exploited, but as demonstrated below,
it is cheaper than the steps to obtain $\mathbf{H}_{\text{SS}}$ even for point groups of high order.

\section{Results and Discussion}\label{sec:results}

The relativistic exact-decoupling approaches with the consideration of minimum
computational requirement discussed in Section~\ref{sec:formalism}
have been implemented into the {\sc Turbomole} \cite{turbomole} program package.
The implementation includes both two-component and scalar versions.
Point group symmetries are available for scalar-relativistic calculations,
while the double point group and time reversal symmetries
have not yet been implemented for the two-component case.
For local relativistic approximations, both DLH and DLU can be used,
although DLU for all atoms is the default local scheme.
Since a finite nucleus model has not been implemented yet,
the point charge nuclei are used throughout. The speed of light was set
to 137.0359895 atomic units.

\subsection{Basis sets}

The exact decoupling approaches require all-electron basis sets, which in particular
for heavier elements may significantly differ from non-relativistic sets.
For the desired application involving silver clusters of more than 300 atoms we decided
to derive an appropriate double-zeta basis set from a non-relativistic all-electron basis optimized
for the d$^{10}$s$^1$ state some years ago, SVPalls1 \cite{MayAhlrichs},
with the contraction pattern \{633311/5331/53\}.
For better comparability with the results obtained with a double-zeta basis used in combination
with a Dirac--Hartree--Fock (dhf) effective core potential (ECP) \cite{Figgen}, dhf-SV(P) \cite{WeigendBaldes},
the contraction pattern for the d shells was changed to \{5211\}.
Exponents and contraction coefficients were optimized simultaneously by repeated calculation
of numerical gradients of the restricted open-shell X2C/Hartree--Fock total energy followed
by a relaxation procedure. The most diffuse p-function which corresponds to
the unoccupied 5p orbital was kept fixed.
For the two-component calculations a (1p1d)/[4p1d] \{-/4/1\} set was added, which was optimized
in the same manner, but by minimizing the total Hartree--Fock energy obtained within the
two-component formalism. 
Resolution-of-the-identity (RI) density-fitting techniques \cite{WeigendKattannekAhlrichs} dramatically improve the efficiency of
two-electron integral calculations, but require sets of auxiliary functions fitting the
products of (orbital) basis functions.
Typical errors of carefully preoptimized sets amount to several ten $\mu$E$_{\text{H}}$ in atomic calculations.
For molecules, RI errors are smaller by about one order of magnitude than those of orbital basis sets.
This accuracy is achieved for the auxiliary basis fitted for the Ag double-zeta orbital bases,
when completely decontracting the first s function.
The original auxiliary basis is available from the internet \cite{BaldesBS}, it is constructed
from the def2-SVP auxiliary basis \cite{Weigend} by adding a (1s)/[18s] set.
The orbital basis (rSV(P)alls1), the patch for two-component calculations
(entire basis termed rSV(P)-2c), and the auxiliary basis are available as supplementary material \cite{suppinf},
as well as two large even-tempered basis sets, one to be used in connection with the
dhf-ECP,(16s16p13d4f2g), and one for the all-electron calculations, (43s31p20d4f2g).

\subsection{Numerical comparison of computation times}

The computation times on the Intel Xeon E5430 central processing unit (CPU) (serial version, one core)
for the different relativistic
decoupling approaches are compared at the example of the Ag$_{13}$ cluster.
119 primitives were contracted to 46 basis functions for each Ag atom.
The dimension of the matrices involved in the setup of the relativistic one-electron Hamiltonian
is then roughly three times the dimension of the matrices used in the SCF iterations such as the Fock matrix.
For the two-component calculation, both dimensions are doubled
if no symmetries were considered.
To illustrate the relative cost of the relativistic transformation
in a whole calculation,
two types of SCF calculations were performed.
One is a standard Hartree--Fock calculation, the other one is DFT
(with the BP86 functional \cite{Becke1988,Perdew1986})
in combination with the resolution of the identity approximation.

%---------------------------------------------------------------------------------------------
\begin{table}[H]
\caption{Computation times (in seconds) of Ag$_{13}$ for the evaluation of relativistic Hamiltonians
and for one SCF iteration within a scalar-relativistic approach.
HF-SCF and DFT-SCF denote a single SCF iteration of a Hartree--Fock and DFT/RI calculation, respectively.
46 contracted basis functions from 119 primitives were employed for each Ag atom.}
\label{tab_cputime12c}
%\begin{ruledtabular}
\begin{center}
\begin{tabular}{crrrrr}
\hline\hline
Spin-Orbit   &no    &no    &no   &yes    &yes   \\
Symmetry     &$C_1$ &$C_1$ &$I_h$&$C_1$  &$C_1$ \\
Local Approximation&none      &DLU   & none    &   none    &DLU   \\
\hline
DKH2         & 11.3 &  1.3 & 1.4 & 178.0 & 13.2 \\
DKH3         & 14.5 &  1.3 & 1.6 & 358.0 & 14.0 \\
DKH4         & 19.8 &  1.4 & 2.3 & 669.8 & 15.3 \\
DKH5         & 28.2 &  1.6 & 3.2 &1169.5 & 12.9 \\
DKH6         & 38.4 &  1.8 & 4.4 &1713.1 & 15.2 \\
DKH7         & 51.2 &  2.1 & 5.9 &2456.8 & 22.9 \\
DKH8         & 69.7 &  2.4 & 7.9 &3456.2 & 27.3 \\
DKH9         & 95.3 &  2.9 &10.7 &4901.7 & 33.5 \\
X2C          & 69.2 &  2.2 & 5.6 &1707.5 & 21.0 \\
BSS          & 71.0 &  2.2 & 5.8 &1861.3 & 21.6 \\
\hline
HF-SCF       &151.0 &151.0 & 3.2 & 310.9 &310.9 \\
DFT-SCF      &  5.7 &  5.7 & 0.2 &  22.4 & 22.4 \\
\hline\hline
\end{tabular}
\end{center}
%\end{ruledtabular}
\end{table}
%---------------------------------------------------------------------------------------------

The computation times for the evaluation of relativistic Hamiltonians
and for one SCF iteration step are presented in Table~\ref{tab_cputime12c}.
We see that the computation of the X2C Hamiltonian is slightly faster than
the BSS Hamiltonian since three additional matrix multiplications are required for
the BSS approach as is evident from Table~\ref{tab_matrixops}.
For scalar calculations,
the computation time of the X2C Hamiltonian is very close to that of DKH8, which has been observed before \cite{peng12_tca}.
The fastest DKH2 approach is about five times faster than that of the X2C approach (for the setup of the one-electron Hamiltonian).
Comparing to the computation times of SCF iterations, one Hartree--Fock iteration is about
twice as expensive as the X2C transformation.
Because several tens of iterations are usually required to obtain converged
results, the SCF iterations dominate the whole computation time in the Hartree--Fock case,
while the RI-DFT calculations are much faster (only 5.7
seconds for one SCF iteration).
The calculation of the relativistic one-electron Hamiltonian thus becomes dominating in the DFT case.
However, the computation time can be dramatically reduced by the DLU approximation.
The point group symmetry can be exploited in scalar-relativistic calculations, and
$I_h$ is the maximum symmetry.
As one can see from Table~\ref{tab_cputime12c},
computation times for both the relativistic transformation and the SCF
are then reduced by almost the same factor.

In the two-component case, the calculation of the relativistic one-electron Hamiltonian becomes much slower than in the scalar case.
The formal ratio of two-component to scalar-relativistic transformation is 32.
The doubled dimension of matrices contributes a factor of eight and
the multiplication of complex numbers contributes another factor of four.
The actual ratios for different relativistic approaches resemble the formal ratio.
The computation time of the X2C Hamiltonian is very close to that of DKH6.
But DKH2 is ten times faster than X2C in this two-component case.
The computational cost for the calculation of the relativistic one-electron Hamiltonian shows a
dramatic increase from scalar to two-component,
while the SCF time is only slightly increased.
This is due to several factors.
The required primitive repulsion integrals are the same as in the SCF stage,
the electron density is always real, and
time-reversal symmetry had been used in the two-component SCF calculations.
The two-component relativistic Hamiltonian construction is now the bottleneck of the whole calculation.
For example, the evaluation of the two-component X2C Hamiltonian requires about 100
times more computation time than one DFT-SCF iteration.
One therefore has to make use of the local relativistic approximation.
If the DLU approximation is employed, the computation time
of the relativistic transformation can be reduced
roughly by a factor of $1/L^2$ where $L$ denotes the number of atoms.
The construction of the relativistic one-electron Hamiltonian is then no longer the bottleneck.
However, the DKH2 approach can be used even without the DLU approximation,
since its computation time is comparable to the SCF iteration in both the scalar
and the two-component case.

\subsection{Application: Cohesive energy of Ag$_L$ Clusters}

\begin{figure}[H]
\caption{\label{structures}Pictorial presentation of octahedral (left) and icosahedral (right) Ag$_L$ cluster structures with $L$=13, 55, 147, and 309.}
\begin{center}
\includegraphics[scale=.55]{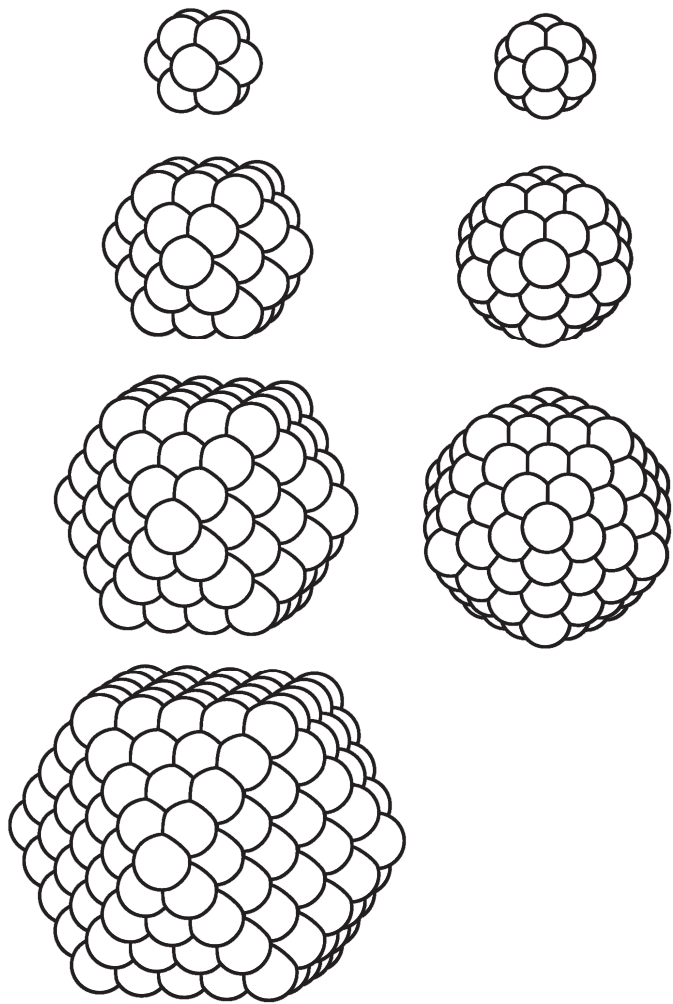}
\end{center}
\end{figure}

We begin this section with the assessment of the quality of the basis sets in atomic DFT calculations
(BP86, medium gridsize (m3)). Then, the influence of SO coupling, density functional, size of the numerical grid in DFT,
basis set superposition error, and of Jahn--Teller distortions is studied for octahedral Ag$_{13}$.
Finally, calculations of octahedral and icosahedral Ag$_L$ clusters with $L$=55, 147, and 309 atoms are presented (see Fig.\ \ref{structures} for a pictorial presentation of these silver clusters).
Calculated cohesive energies of the clusters are extrapolated to the bulk value and the performance of the
different techniques (full X2C with/without symmetry exploitation, local X2C) is demonstrated and discussed.

%---------------------------------------------------------------------------------------------
\begin{table}[H]
\caption{Basis set errors for the electronic BP86 energy of the Ag atom (gridsize m3) in E$_{\text{H}}$.
For details concerning bases and RI-J auxiliary bases see text and supplementary material.
In case of the all-electron calculations '1c' denotes one-component X2C and
$\Delta$(2c) the difference of one- and two-component X2C energies.
In the case of the ECP-based calculations, 'scalar-relativistic' refers to 
one-component ECPs, $\Delta$(2c) denotes the difference from one- to two-component ECPs
with dhf-SV(P)-2c bases used in both calculations. 'basis1' denotes the reference
basis set 43s31p20d4f2g, whereas 'basis2' refers to the reference basis set 16s16p13d4f2g.
}
\label{tab:res1}
\begin{center}
\begin{tabular}{lrrrrrr}
\hline\hline
               & \multicolumn{4}{c}{{all-electron}}    & \multicolumn{2}{c}{{dhf-ECP(28)}} \\
               & basis1       & SV(P)alls1   & rSV(P)alls1   & rSV(P)alls1-2c & basis2      & dhf-SV(P) \\
\hline
non-rel.            & -5200.587052 & -5199.872643 & -5108.671268  &  -5111.223910  &        ---  &         ---   \\
$\Delta$RI-J   &     0.000009 &     0.000009 &     0.000068  &      0.000067  &        ---  &         ---   \\
1c             & -5316.242277 & -5280.524199 & -5315.547762  &  -5315.564554  & -147.041728 & -147.023711   \\
$\Delta$RI-J   &     0.000069 &     0.000008 &     0.000074  &      0.000074  &    0.000007 &    0.000055   \\
$\Delta$(2c)   &    -0.612892 &    +0.371194 &    +0.466913  &     -0.509039  &   -0.039243 &   -0.034318   \\
\hline\hline
\end{tabular}
\end{center}
\end{table}
%---------------------------------------------------------------------------------------------

Data for atomic DFT(BP86) calculations are collected in Table~\ref{tab:res1}.
Errors, i.e., differences to the reference basis, for rSV(P)alls1 in scalar-relativistic calculations are
similar to those of SVPalls1 in non-relativistic calculations (0.695 vs.\ 0.714 E$_{\text{H}}$).
By contrast, errors for SV(P)alls1 in relativistic calculations amount to about 36 E$_{\text{H}}$, which is too much,
even if most of this cancels for bond energies; a similar observation holds true for rSV(P)alls1
used for non-relativistic calculations.
For a qualitatively correct description of energy changes by SO coupling in self-consistent
two-component calculations the (1p1d)/[4p1d] is required, otherwise even the sign is incorrect.
With this patch the reference result for the energy change by SO coupling (-0.613 E$_{\text{H}}$)
is quite well reproduced (-0.509 E$_{\text{H}}$).
In the ECP case, the lowering is, of course, much smaller, as those
shells which are most strongly affected by SO coupling are included in the ECP.
The result for the reference basis (-0.039 E$_{\text{H}}$) is well reproduced by
the dhf-SV(P)-2c basis (-0.034 E$_{\text{H}}$).
Furthermore, the RI-J auxiliary basis described above yields very reasonable results;
the largest errors amount to 70 $\mu$E$_{\text{H}}$,
which is several orders below the basis set error.
Next, we investigate the influence of SO coupling, functional, gridsize, and
of Jahn--Teller distortion on the cohesive energy of octahedral Ag$_{13}$, and determine the basis set
superposition error of the respective bases for this system.
%---------------------------------------------------------------------------------------------
\begin{table}[H]
\caption{Cohesive energies (atomization energies per atom), in kJ/mol, obtained for octahedral Ag$_{13}$ with different methods. 
Basis sets are of type rSV(P)alls1 (this work, see text) for X2C and of type dhf-SV(P) \cite{WeigendBaldes} for the ECP calculations. 
The first row 'standard' denotes the cohesive energy obtained for $O_h$ symmetry with the BP86 density functional in a restricted Kohn-Sham calculation 
(the HOMO, $t_{2g}$, is occupied with five electrons). The RI approximation is used for the Coulomb part (see text for the construction of the auxiliary basis), 
for the DFT part coarse grids (grid1) are used during iterations and a medium grid (grid3) for the final energy calculation (grid m3 \cite{treutler}). 
The rows below the first one list energy changes upon changing each parameter separately and keeping the others as defined in row 'standard'. 
Reference bases (second row) are even-tempered bases of size (16s16p13d4f2g) for column 'ECP' and (43s31p20d4f2g) for column 'X2C'; for details of the 
counterpoise correction (third row) see text. 'TPSS' denotes the TPSS density functional \cite {tpss} employed instead of BP86, 
'noRI' refers to the exact calculation of the Coulomb part via four-center integrals (instead of invoking the RI approximation). 
Rows 'grid3' and 'grid6' concern modifications of the integration grid for the DFT part: 'grid3' ('grid6') denote the employment of the medium (fine) 
grids throughout. In the next two rows structure constraints are relaxed to $D_{4h}$ and $D_{3d}$ symmetry, respectively. The data in row '2c' are 
obtained with basis sets extended for two-component calculations, dhf-SV(P)-2c \cite {WeigendBaldes} and rSV(P)alls1-2c (this work, see text).
\label{tab:res2}}
\begin{center}
\begin{tabular}{lrr}
\hline\hline
Method     & ECP  & X2C \\
\hline
standard         &    144.53  & 162.45    \\
reference basis  &     +7.59  & -15.08    \\
counter-poise corrected   &     -4.06  & -22.76    \\
TPSS             &    +11.47  & +11.25    \\
noRI            &     +0.04  &  -0.05    \\
grid3            &      0.00  &   0.00    \\
grid6            &     -0.07  &  -0.10    \\
$D_{4h}$         &     +0.59  &  +0.55    \\
$D_{3d}$         &     +0.48  &  +0.37    \\
2c               &     +3.74  &  +5.19    \\
\hline\hline
\end{tabular}
\end{center}
\end{table}
%---------------------------------------------------------------------------------------------

The data shown in Table~\ref{tab:res2} reveal the following picture.
At first glance, X2C yields a significantly higher cohesive energy than the
ECP calculation (+18 kJ/mol), but, when employing the respective reference bases,
the calculated cohesive energy becomes
larger by 7 kJ/mol for the ECP case and smaller by 15 kJ/mol for the X2C case,
so that the two methods yield very similar results with the reference basis.
The reason for this is a comparably large basis set superposition error (BSSE)
of about 23 kJ/mol in case of the rSV(P)alls1 basis: for the central atom the
presence of the basis functions of the
outer atoms leads to an energy lowering by 14.6 mE$_{\text{H}}$, while for a shell atom
the respective lowering is 8.2 mE$_{\text{H}}$.
For the dhf-SV(P) basis these effects are much smaller, 2.0 and 1.5 mE$_{\text{H}}$,
respectively, which leads to a correction of about 4 kJ/mol for the cohesive energy.
We note that this is not a particular problem of the rSV(P)alls1 basis:
the errors of the non-relativistic SVPalls1 basis (calculated for the
non-relativistic case) are even larger: 34.2 and 22.2 mE$_{\text{H}}$, respectively.
The BSSE for the reference bases amounts to about 1 kJ/mol for the all-electron
calculation and to about 2 kJ/mol for the ECP calculation.
Employing the TPSS functional \cite {tpss} instead of BP86 leads to 11 kJ/mol higher
cohesive energies for both X2C and ECP. The finer DFT grids
do not change this quantity significantly ($<$0.1 kJ/mol).
The influence of the Jahn--Teller distortion due to the partially filled
highest-occupied molecular orbital (HOMO) also is rather small (about 0.5 kJ/mol). SO coupling yields an increase
of values by 3.7 kJ/mol for the ECP calculation and of
5.2 kJ/mol for the X2C case (for the extended bases dhf-SV(P)-2c and rSV(P)alls1-2c).

%---------------------------------------------------------------------------------------------
\begin{table}[H]
\caption{Cohesive energies (atomization energies per Ag atom) for icosahedral and octahedral Ag$_L$ clusters in kJ/mol.
The extrapolated values are obtained from a linear regression of the
cohesive energy versus $L^{-1/3}$. 'pg' denotes the point group symmetry of the clusters and 'cp-corrected' the counterpoise correction.}
\label{tab:res3}
\begin{center}
\begin{tabular}{lccccc}
\hline\hline
        & &\multicolumn{2}{c}{{BSSE-affected}} & \multicolumn{2}{c}{{cp-corrected}} \\
$L$     & pg & ECP & X2C     & ECP & X2C                    \\
\hline
13      &    $I_h$     & 143       & 158           & 139       & 135                          \\
13      &    $O_h$     & 145       & 163           & 144       & 140                          \\
55      &    $I_h$     & 189       & 215           & 185       & 189                          \\
55      &    $O_h$     & 186       & 210           & 182       & 184                          \\
147     &    $I_h$     & 202       & 230           & 198       & 202                          \\
147     &    $O_h$     & 199       & 226           & 195       & 198                          \\
309     &    $O_h$     & 206       & 236           & 201       & 207                          \\
\hline
extrap  &              & 245       & 282           & 239       & 250                          \\
Exp.\ \cite{ecohag} &  &           &               &  \multicolumn{2}{c}{285}                 \\
\hline\hline
\end{tabular}
\end{center}
\end{table}
%---------------------------------------------------------------------------------------------

dhf-ECP/dhf-SV(P)/BP86/RI and X2C/rSV(P)alls1/BP86/RI calculations were
carried out for octahedral and icosahedral Ag clusters of 13, 55, 147, and 309 atoms
in order to investigate the quality of
the prediction for the bond energy obtained with ECP and X2C, respectively,(see Table \ref{tab:res3}) and
to assess the computational effort.
Data obtained with X2C are larger by 15-30 kJ/mol, if not corrected for BSSE;
when the counterpoise correction is considered (by subtracting the error data obtained for Ag$_{13}$
for each inner and surface atom),
cohesive energies obtained with the two methods become very similar. Differences
typically amount to 4 kJ/mol with a slight increase from smaller to larger clusters.
Note that this is smaller than, e.g., the difference due to the choice of the density functional, as measured by a comparison of TPSS and BP86 results.
Both methods predict only small differences between icosahedral and octahedral species;
the indicated slight preferences are the same for both methods: $O_h$ for Ag$_{13}$, $I_h$ else.
From these data the cohesive energy of the bulk may be obtained by a linear regression
of the cohesive energies calculated for the $L$-atomic clusters versus $L^{-1/3}$.
It amounts to 239 kJ/mol for the ECP calculation and to 250 kJ/mol for the X2C calculation.
The agreement with the experimental value of 285 kJ/mol \cite{ecohag} is reasonable and can be improved
by correcting it with the data obtained for Ag$_{13}$.
We add the difference of the BSSE corrected numbers to that
obtained for the reference basis, i.e., 7.7 kJ/mol (X2C) and 11.7 kJ/mol (ECP),
and the effect of SO coupling, 3.7 kJ/mol (ECP) and 5.2 kJ/mol (X2C),
thus yielding 254 kJ/mol (ECP) and 263 kJ/mol (X2C).
If one further considers data for TPSS by adding the difference between TPSS and BP86,
one finds 257 kJ/mol (ECP) and 274 kJ/mol (X2C), which is close to the experimental value of 285 kJ/mol \cite{ecohag},
in particular in case of X2C.
Overall, for the present case both techniques yield results of similar accuracy. Generally, from X2C one may expect greater reliability, as it is a well-founded theoretical improvement.

%---------------------------------------------------------------------------------------------
\begin{table}[H]
\caption{Computation times in seconds (on an Intel Xeon X3460 CPU with 2.8 GHz) of Ag$_L$ clusters
for different variants of the X2C procedure and for the
subsequent SCF loop. The subscript 'all' denotes the full X2C procedure, i.e. without symmetry
in the X2C step, the subscript 'sym' denotes the full procedure with symmetry blocking in the X2C step, and 'loc' refers to the
local DLU scheme (without exploiting point-group symmetry in the set-up of the X2C/DLU one-electron Hamiltonian) with the resulting error in the cohesive energy in kJ/mol.
$n_{\text{iter}}$ is the number of iterations needed for SCF convergence in ECP ('SCF@ECP')and X2C ('SCF@X2C') calculations. 'pg' denotes the point group symmetry of the cluster.
'SO' refers to spin-orbit coupling and 'UKS' to (scalar-relativistic) unrestricted Kohn--Sham.}
\label{tab:res4}
\begin{center}
\begin{tabular}{rrrrrrrrrr}
\hline\hline
$L$  & pg & X2C$_{\rm all}$ & X2C$_{\rm sym}$ & X2C$_{\rm loc}$ & err(loc) & SCF@X2C & $n_{\text{iter}}$ & SCF@ECP & $n_{\text{iter}}$ \\
\hline
 13  & $I_h$     &    66  &     7  &   2 &  0.13  &       7 &  29 &      4 &  12  \\
 13  & $O_h$     &    65  &     6  &   2 &  0.14  &       8 &  89 &      8 &  78  \\
 55  & $I_h$     &  5639  &   457  &  42 &  0.57  &     946 & 124 &    464 & 113  \\
 55  & $O_h$     &  6016  &   385  &  43 &  0.53  &    1035 & 122 &    308 & 115  \\
 147 & $I_h$     & 97773  &  7603  & 457 &  0.60  &   21966 & 248 &   9865 & 250  \\
 147 & $O_h$     &   ---  &  6791  & 409 &  0.92  &   21086 & 273 &   8243 & 259  \\
 309 & $O_h$     &   ---  & 51911  & 1875&  0.14  &  101377 & 297 &  61967 & 459 \\
 13  & $C_1$/UKS &   108  &   ---  &   2 &  0.20  &    2271 & 198 &    697 & 212  \\
 13  & $C_1$/SO  &  1642  &   ---  &  20 &  0.20  &    3581 & 160 &    236 &  21  \\
\hline\hline
\end{tabular}
\end{center}
\end{table}
%---------------------------------------------------------------------------------------------

We finally discuss the computational effort for the X2C calculation. As noted above,
one goal of this work is to accomplish an implementation, for which the X2C part does not dominate the computation time.
As shown in Table~\ref{tab:res4}, this goal is reached for the calculations without symmetry
exploitation for both the X2C and the two-electron part,
calculations exploiting symmetry in both parts, and for the DLU approximation.
The latter reduces the CPU time for X2C to few per cent of the total time,
but also for full X2C typically only 1/4 to 1/3 of the time is spent in the X2C part,
as long as symmetry is exploited.
This is remarkable, as the number of primitives, which determines the dimension of the matrices in the X2C transformation,
is nearly three times as large as the number of contracted basis functions,
e.g., 31518 primitives versus 11742 contracted basis functions for Ag$_{309}$.
Admittedly, the number of iterations is larger than usual, due to the metallic structure
and the chosen parameters for the Fock matrix update
('fermi smearing' \cite{NavaSierkaAhlrichs} and high damping). Nevertheless,
this ratio is not untypical.
The chosen procedure, BP86 functional with grid m3 and RI-J is one of the fastest methods
to calculate two-electron interactions.
For meta-generalized-gradient-approximation density functionals or finer grids one might expect an increase in time by a factor of at least two,
and, as soon as Hartree--Fock exchange is involved, by a factor of about ten.
Moreover, the application of difference densities in
the SCF iteration leads to much less than a linear increase of computation costs
with the number of iterations.
For two-component calculations a similar ratio is obtained.
Compared to the ECP approximation, the effort for X2C, of course, is higher,
because of the larger basis and the higher effort for the one-electron integrals
(which for ECPs is almost negligible), but it is not that much.
The ratio of CPU times for DLU-X2C and ECP typically is somewhat above two and
for full X2C (with symmetry blocking) about three.
For larger bases these ratios will be even smaller.

Interestingly, we found that the errors introduced by the DLU approximation are by one
or two orders of magnitude larger than what we observed in Ref.\ \cite{peng12_jcp}.
For example, the DLU-X2C error of Pb$_9^{2+}$ is 0.05 mE$_{\text{H}}$
which amounts to an error of 0.015 kJ/mol in cohesive energy,
while the DLU-X2C error of Ag$_{13}$ computed in this article is 0.20 kJ/mol.
The error is more than ten times larger for molecules of almost the same size, though still negligibly small considering the accuracy of the electronic structure method employed.  
The main difference of those two calculations are the type and size of the basis functions. 
Ref. \cite{peng12_jcp} employed the general contracted 
ANO \cite{roos05_jpca,roos05_cpl,roos08_jpca}
basis functions of double-zeta quality, 
while segmented contracted basis functions together with single diffuse 
functions are used in this work.
We found that the DLU error mainly stems from the diffuse function, which does not come as a surprise since our local (atomic) construction of the unitary matrix is defined by the atom-centered basis functions. 
To illustrate their effect, we removed the most diffuse functions of all orbital angular momentum (s,p,d) in the unrestricted Kohn--Sham calculation of the Ag$_{13}$ cluster. 
As can be seen from Table~\ref{tab:res5},
the error is then reduced from 0.20 kJ/mol to 0.09 kJ/mol. 
If a set of more diffuse functions (with exponents 1/3 of the smallest one) are added to the basis set,
the DLU error increases to 0.39 kJ/mol.
%---------------------------------------------------------------------------------------------
\begin{table}[H]
\caption{DLU errors of the cohesive energy in kJ/mol for the UKS calculation of 
Ag$_{13}$ with diffuse basis functions removed or added (as indicated by the sign).} 
\label{tab:res5}
\begin{center}
\begin{tabular}{ccccc}
\hline\hline
basis          & -(1s1p1d) & ~~~~~~0~~~~~& +(1s1p1d) & +(2s2p2d) \\
\hline
err(loc) &   0.09    & 0.20        &   0.39    &  0.40     \\
\hline\hline
\end{tabular}
\end{center}
\end{table}
%---------------------------------------------------------------------------------------------
Adding more diffuse functions does not change the DLU error much (0.39 kJ/mol becomes 0.40 kJ/mol) because they contribute hardly to occupied molecular orbitals and total electronic energies.
It is obvious that a diffuse function largely affects the DLU error as a
diffuse function has a long tail, so that it has significant contribution to the basis space of its neighboring atoms.  
The DLU approximation only considers AO functions of the same atom as a local 
block, it neglects the contributions from diffuse function of other atoms. 
We also considered examples of Ref. \cite{peng12_jcp} using the ANO basis in fully uncontracted form.
The DLU cohesive energy error of I$_5^+$ is 0.18 kJ/mol,
while the error on the total electronic energy reported in Ref. \cite{peng12_jcp} is 0.007 kJ/mol.
If we remove the most diffuse functions of all orbital angular momentum 
from this uncontracted set, the DLU error decreases to 0.047 kJ/mol.

\section{Conclusions}\label{sec:conclusion}

Relativistic exact decoupling can be achieved by either
the X2C approach or the BSS approach. The X2C approach is recommended since it is 
simpler as it avoids the additional free-particle Foldy--Wouthuysen transformation of BSS, which is
an atavism of the sequential DKH decoupling approach.
With the DKH method, results as accurate as in the exact-decoupling
approaches can be obtained by choosing an appropriate (high) expansion order.
Such order depends on what atoms and what type of properties
are considered.

We have implemented and thoroughly analyzed these relativistic two-component exact-decoupling methods
in the quantum chemistry package {\sc Turbomole}
(scalar-relativistic methods are also available).
The formulae and algorithms of the relativistic Hamiltonians implemented
have been carefully organized so that their calculation can be performed most efficiently.
Parallelization techniques and, especially, point group symmetries can be exploited to accelerate the calculation.
Furthermore, we have elaborated further on our local relativistic approximation,
i.e., the DLU approach, which constructs the unitary transformation form diagonal (atomic) blocks only in order to 
significantly reduce the computational cost.
If the atoms involved are moderately heavy, one can safely use Hess' standard DKH2 Hamiltonian,
which is five to ten times faster to evaluate than the X2C Hamiltonian.

With this efficient implementation, the exact-decoupling approach
can be expected to become the standard relativistic all-electron approach in the future.
However, there are still some issues that need to be worked out.
For instance, we might further develop the DLU approximation to reduce its error by 
employing a more rigorous definition of an atomic block in the decomposition of the unitary transformation than
the one which is provided by the atom-centered basis functions of a given basis set. One might even use
the reduction of the DLU error in the fitting of relativistic basis sets which brings us to the next task, namely
the development of all-electron basis sets suitable for the exact-decoupling
approach. These developments are in progress in our laboratory.

\begin{acknowledgements}

This work has been supported by the Swiss National Science Foundation SNF.

\end{acknowledgements}

\appendix

\section{}\label{secapp}

The $\bm{X}$ matrix in the original non-orthogonal basis representation reads
\begin{eqnarray}\label{defxmat}
\bm{X}=\bm{C}_{S}^{+}\left(\bm{C}_{L}^{+}\right)^{-1},
\end{eqnarray}
where $\bm{C}_{L}^{+}$ and $\bm{C}_{S}^{+}$ are eigenvector
coefficients of Eq.~(\ref{moddiraceq}).
Since the orthonormal basis representation is obtained by applying the transformation
$\mathbb{K}$ defined in Eq.~(\ref{defk}),
the eigenfunctions of those two basis representations are connected by
\begin{eqnarray}
\bm{C}_{L}^{+}&=&\bm{K}\bm{C}_{L'}^{+},\label{defclortho}\\
\bm{C}_{S}^{+}&=&2c\bm{K}\bm{p}^{-1}\bm{C}_{S'}^{+}\label{defcsortho}.
\end{eqnarray}
According to the definition of the $\bm{X}'$ matrix in Eq.~(\ref{defxortho}),
we obtain its relation to the $\bm{X}$ matrix by inserting Eq.~(\ref{defclortho})
and (\ref{defcsortho}) into Eq.~(\ref{defxmat})
\begin{eqnarray}\label{relationxx}
\bm{X}=2c\bm{K}\bm{p}^{-1}\bm{X}'\bm{K}^{-1}.
\end{eqnarray}
Then, we investigate the $\bm{X}$-dependent term of $\bm{U}_{\rm X2C}^{L}$
in Eq.~(\ref{ulnonortho})
\begin{eqnarray}
\frac{1}{2c^2}\bm{X}^{\dag}\bm{T}\bm{X}&=&
2(\bm{K}^{\dag})^{-1}\bm{X}'^{\dag}\bm{p}^{-1}\bm{K}^{\dag}\bm{T}\bm{K}\bm{p}^{-1}\bm{X}'\bm{K}^{-1}\\
&=&(\bm{K}^{\dag})^{-1}\bm{X}'^{\dag}\bm{X}'\bm{K}^{-1},\label{condxtx}
\end{eqnarray}
where we have used
\begin{eqnarray}
\bm{K}^{\dag}\bm{T}\bm{K}=\bm{p}^2/2.
\end{eqnarray}
According to the properties of the $\bm{K}$ matrix
\begin{eqnarray}
\bm{K}^{\dag}\bm{S}\bm{K}=\bm{K}^{\dag}\bm{S}^{\frac{1}{2}}\bm{S}^{\frac{1}{2}}\bm{K}=\bm{I},
\end{eqnarray}
one can see that $\bm{S}^{\frac{1}{2}}\bm{K}$ is a unitary matrix
(noting that $\bm{S}^{\frac{1}{2}}$ is hermitian).
Denoting the unitary matrix as $\bm{U}$, we have
\begin{eqnarray}
\bm{S}^{\frac{1}{2}}\bm{K}=\bm{U}
&\Longrightarrow& \bm{K}=\bm{S}^{-\frac{1}{2}}\bm{U}, \label{conduk} \\
&\Longrightarrow& \bm{K}^{-1}=\bm{U}^{\dag}\bm{S}^{\frac{1}{2}}, \label{condukinv} \\
&\Longrightarrow& (\bm{K}^{\dag})^{-1}=\bm{S}^{\frac{1}{2}}\bm{U}. \label{condukdinv}
\end{eqnarray}

Now, we can derive the equivalence of Eq.~(\ref{ulnonortho}) and Eq.~(\ref{ulortho})
step by step as follows
\begin{eqnarray}
\bm{U}_{\rm X2C}^{L}&=&\bm{S}^{-\frac{1}{2}}\left(\bm{S}^{-\frac{1}{2}}
(\bm{S}+\frac{1}{2c^2}\bm{X}^{\dag}\bm{T}\bm{X})
\bm{S}^{-\frac{1}{2}}\right)^{-\frac{1}{2}}\bm{S}^{\frac{1}{2}} \label{step1} \\
&=&\bm{S}^{-\frac{1}{2}}\left(\bm{I}+\bm{S}^{-\frac{1}{2}}
(\bm{K}^{\dag})^{-1}\bm{X}'^{\dag}\bm{X}'\bm{K}^{-1}
\bm{S}^{-\frac{1}{2}}\right)^{-\frac{1}{2}}\bm{S}^{\frac{1}{2}} \label{step2} \\
&=&\bm{S}^{-\frac{1}{2}}\left(\bm{I}+
\bm{U}\bm{X}'^{\dag}\bm{X}'\bm{U}^{\dag}
\right)^{-\frac{1}{2}}\bm{S}^{\frac{1}{2}} \label{step3} \\
&=&\bm{S}^{-\frac{1}{2}}\left(\bm{U}(\bm{I}+
\bm{X}'^{\dag}\bm{X}')\bm{U}^{\dag}
\right)^{-\frac{1}{2}}\bm{S}^{\frac{1}{2}} \label{step4} \\
&=&\bm{S}^{-\frac{1}{2}}\bm{U}\left(\bm{I}+
\bm{X}'^{\dag}\bm{X}'\right)^{-\frac{1}{2}}
\bm{U}^{\dag}\bm{S}^{\frac{1}{2}} \label{step5} \\
&=&\bm{K}\left(\bm{I}+
\bm{X}'^{\dag}\bm{X}'\right)^{-\frac{1}{2}}
\bm{K}^{-1}=\bm{K}\bm{R}'\bm{K}^{-1}. \label{step6}
\end{eqnarray}
Eq.~(\ref{condxtx}) was inserted into (\ref{step1}) to get (\ref{step2}).
From (\ref{step2}) to (\ref{step3}), the various forms of the $\bm{K}$ matrix
in Eq.~(\ref{condukinv}) and (\ref{condukdinv}) were used.
The $\bm{S}^{-\frac{1}{2}}$ matrix under the bracket is then eliminated.
The identity $\bm{I}=\bm{U}\bm{U}^{\dag}$ was employed in the next step.
From (\ref{step4}) to (\ref{step5}), the unitary matrices are moved to the
outside of the inverse square root bracket since the following equation
\begin{eqnarray}\label{uauiden}
\left(\bm{U}\bm{A}\bm{U}^{\dag}\right)^{k}=\bm{U}\bm{A}^{k}\bm{U}^{\dag}
\end{eqnarray}
is satisfied for any diagonalizable matrix $\bm{A}$ such as hermitian matrices.
$k$ could be any real number and $k=-\frac{1}{2}$ is of course valid for Eq.~(\ref{uauiden}).
Using again the forms of the $\bm{K}$ matrix in Eq.~(\ref{conduk}) and (\ref{condukinv}),
we obtain the final Eq.~(\ref{step6}), which is exactly the form in orthonormal
basis representation as presented in Eq.~(\ref{ulortho}).

Comparing Eq.~(\ref{ulsnonortho}) with Eq.~(\ref{ulsortho}),
we find that the following identity
\begin{eqnarray}\label{idenxx}
\bm{X}\bm{K}=2c\bm{K}\bm{p}^{-1}\bm{X}'.
\end{eqnarray}
is required to prove the equivalence of
the small component of X2C decoupling transformation matrices.
The above equation can be obtained by multiplying $\bm{K}$ from the right to each side
of Eq.~(\ref{relationxx}).
Therefore, the two expressions in Eq.~(\ref{usnonortho}) and Eq.~(\ref{usortho}) are
also equivalent.

%---------------------------------------------------------------------------------------------
%\bibliographystyle{aip-mod}
%\bibliography{r2c}

%---------------------------------------------------------------------------------------------

\newpage

\section*{SUPPORTING INFORMATION}

{\footnotesize
\renewcommand{\baselinestretch}{0.6}

}

\end{document}